\documentclass{aa}
\usepackage{txfonts}
\usepackage{graphicx}
\usepackage{amsmath}
\usepackage{natbib}
\usepackage{amssymb}
\bibpunct{(}{)}{;}{a}{}{,} 
\begin{document}

\title{The first stars:  a classification of CEMP-no stars}
\titlerunning{CEMP--no stars}

\author{Andr\'e Maeder, Georges Meynet
}
\authorrunning{Maeder and Meynet}

\institute{Geneva Observatory, Geneva University, CH--1290 Sauverny, Switzerland\\
email: andre.maeder@unige.ch,  georges.meynet@unige.ch
}

\date{Received  / Accepted }

\offprints{Andr\'e Maeder}

\abstract
{}
{We propose and apply a new classification for the  CEMP--no stars, which are  "carbon-enhanced metal-poor" stars with no overabundance  of s-elements and with [Fe/H] generally  $\leq$ -2.5.}
{This classification is based on the changes in  abundances    for the elements and isotopes involved in the CNO, Ne--Na, and Mg--Al nuclear cycles. These abundances  
change very much owing to 
successive  back and forth  mixing motions between the He-- and H--burning regions in massive stars (the "source stars"  responsible for the
chemical enrichment of the CEMP--no stars).}  
{The wide variety of the ratios  [C/Fe], $^{12}$C/$^{13}$C, [N/Fe], [O/Fe], [Na/Fe], 
[Mg/Fe], [Al/Fe], [Sr/Fe], and [Ba/Fe], which are the main characteristics making the CEMP--no  and low s stars so peculiar,
is described well in terms of the proposed nucleosynthetic classification.  We note that
  the  [(C+N+O)/Fe] ratios significantly increase for lower values of [Fe/H].  
}
{The classification  of CEMP-no stars and the behavior of [(C+N+O)/Fe]   support  
 the presence, in  the first stellar generations of the Galaxy, of fast-rotating massive stars  experiencing strong mixing and mass loss (spinstars).
  This result has an impact on the early chemical and
spectral evolution of the Galaxy. }

\keywords{stars: abundances -- stars: massive -- stars: Population III -- Galaxy: evolution}
  
\maketitle
 
 \section{Introduction}

Carbon-enhanced metal-poor (CEMP) stars are  old low-mass stars with very low contents in iron   and in other $\alpha$-elements (with atomic mass number A > 24) that also show impressively  large excesses of carbon, as  well as generally   
of nitrogen and oxygen.  This class  of very   interesting objects  that belong to the first stellar generations in the Universe was initially recognized by 
\citet{Beers1992}.  Several  subclasses of CEMP
stars were distinguished and received an appropriate nomenclature in a review of the very metal-poor stars in the Galaxy \citep{Beers2005}. There are the so--called CEMP--s, CEMP--r/s, CEMP--r,  and CEMP--no stars based  on the contents in s-- (particularly Ba)  and r--elements (particularly Eu), or on  the relative absence of these elements in the case of CEMP--no stars. For recent catalogs, the reader may refer  to \citet{Masseron2010}, \citet{Allen2012}, \citet{Norris4}, and \citet{Hansen2015}. 

The  CEMP--no stars dominate in the range of the lowest [Fe/H] ratios, below [Fe/H] $\approx -3.0$  \citep{Norris4}.  Some stars with low 
s content are sometimes called "low-s stars" and may be related to the subclass of CEMP--no stars. Since  several CEMP-no stars are MS or subgiant stars, their very peculiar CNO contents are likely not due to self-enrichment, but may result from the yields of previous objects
called the "source stars", which are supposed to be among the first stars in the Universe.

What kinds of objects were these first stars? To get an answer, we have to do  careful detective work to understand  the origin
of these very peculiar chemical abundances of CEMP-no stars. This is a fascinating topic that is closely related to nuclear astrophysics.
  Many models have been proposed to try to explain the chemical properties of CEMP-no stars,  such as the discussions by \citet{Nomotoaraa2013}. 
In a recent work  \citep{Maeder2015}, hereafter called Paper I, we provide tests showing that the   CNO abundances  of CEMP-no stars result from the products of 
He-burning (C and O) that have gone through partial mixing and CNO processing in the H-burning shell, before being ejected  to enrich the local interstellar medium.  This is supported by the study of the 
$^{12}$C/$^{13}$C ratios, of the [C/N] and [O/N] ratios. In CEMP-no stars, the elements involved in the CNO cycle cover a  completely different
range of variations than  the $\alpha$-elements (with an atomic mass number higher than  24).  In addition, the heavy elements Na, Mg, and Al  
participating in the  Ne--Na and Mg--Al cycles
behave like the CNO elements and not   like the $\alpha$-elements, which is quite consistent with the operation of H-burning in massive stars.

Our purpose here is to examine some further consequences of the general model of fast-rotating massive stars  (spinstars) with mixing and mass loss  \citep{Meynet2006,Maeder2012}, which we applied to CEMP--no stars  in Paper I. 
    This model 
 accounts for the wide variety of the typical overabundances of C, N, O, Na, Mg, Al, and of s-elements in some cases, as well as for some 
specific relations between these elements (cf. Paper I).
Section 2 proposes detailed classification criteria. In Section 3, we apply this classification to the whole sample of known CEMP-no stars.
Section 4 gives the conclusions.

\section{Criteria for the classification of  CEMP-no stars}

The  classification is  based on the nuclear  processes that have occurred in the source stars and on the possible release by mass loss of the mixture produced  at  various successive steps of mild mixing. If the  C and O elements resulting from He--core burning enter by mild mixing into the H-burning shell,
the nuclear reactions  there  build some new elements, such as $^{13}$C  and $^{14}$N. Now, if these  elements are  again mildly mixed into the He-burning zone, $^{13}$C and $^{14}$N will be destroyed and some  other new elements may be created by $\alpha$ captures,
such as $^{20,  22}$Ne and  $^{24, 25, 26}$Mg,  together with  s-elements of the first peak, such as Sr and Y, and  (depending on the conditions) even of the second peak, 
such as Ba and La.
These products may continue to undergo mixing  and go back to the H-burning region, where  $^{13}$C  and $^{14}$N are built again. Now  
the Ne--Na and Mg--Al
cycles may operate, producing Na and Al. At each step, mass ejection may occur,  preserving the actual  nucleosynthetic products.

In the sequence of  successive back-and-forth motions between the He-- and H--burning  regions, some elements are more or less destroyed, while others are synthesized. This creates the wide variety of the  observed ratios, such as $^{12}$C/$^{13}$C, [C/Fe], [N/Fe], [O/Fe],  [Na/Fe], [Mg/Fe], [Al/Fe], [Sr/Fe], and [Ba/Fe]. These successive stages of mixing have  chemical abundances  different from the case of full or convective mixing, which would lead to the equilibrium abundance ratios that essentially correspond to the highest temperature in the convective region. The processes  of mild mixing produce  large differences between the abundances of the mentioned  elements  and of the $\alpha$-elements with A> 24. Thus, depending on the history of mass loss, very different chemical yields may be produced. In the line of the results of Paper I,  we  establish here the  classification criteria and apply them to individual stellar observations.

\begin{figure}[t]
\begin{center}
\includegraphics[width=7cm, height=5.4cm]{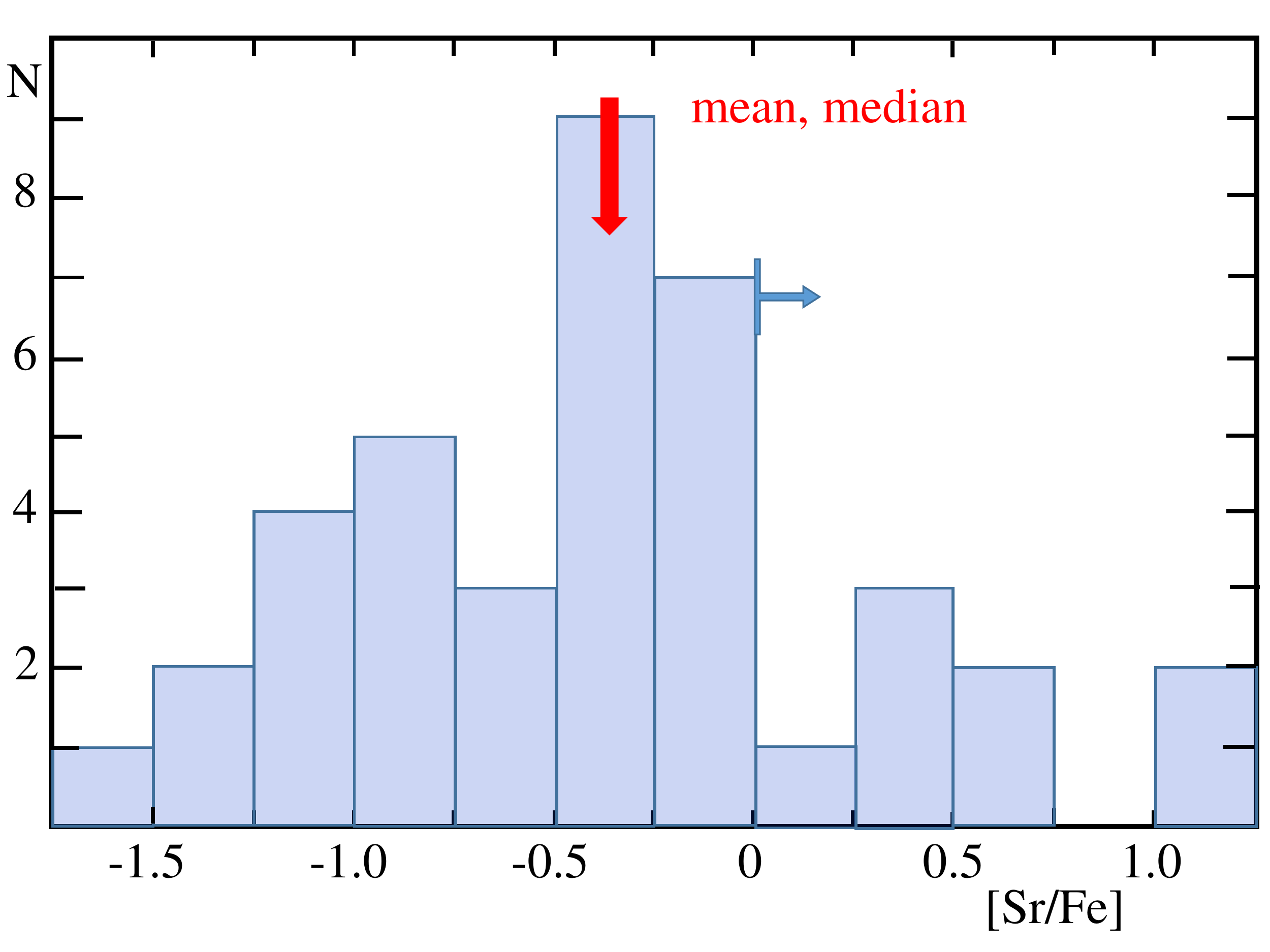}
\caption{Distribution of the ratio [Sr/Fe] for the CEMP-no stars in Table 2. The positions of the mean and median values are indicated. 
The small vertical bar with an arrow on the right indicates above which value we consider that there are significant  excesses. }
\label{SR}
\end{center}
\end{figure}

Table 1 shows the proposed  classification criteria for CEMP-no stars. The first column mentions the main 
physical processes involved, the second one assigns a class or a step in the back-and-forth motions between the He-- and H--burning regions, and the next 
columns  indicate  the possible ranges for the 
 significant abundances ratios defining the considered classes. 
  We note that, on one hand, more subclasses might be needed to fully represent the complex effects of mixing and mass loss, and, on the other, some proposed classes might  not
contain some real objects.  In any case, the comparison of the predicted classes of chemical compositions with the observations will
show us whether the proposed scheme is valid.

\subsection{Classes 0 and 1}
The first kind of possible composition is the initial one, from which the source stars were formed. If these really belonged to the very first stellar generation, their
composition should be the cosmological one of pure hydrogen and helium with tiny amounts of light elements,
  as for lithium with $A(Li)= \log n(Li)/n(H) +12= 2.72$ \citep{Cyburt2008}. These would be extremely metal-poor stars (EMP). In fact, supernovae from massive stars may explode already  after a few million
years, so that the content of the first stellar generations in heavy elements may be slightly different from zero. This initial  composition  would 
correspond to   Class or Step  "0".
Class "0+" is associated to  CNO--burning of a medium with the above initial composition. For CNO reactions to be effective,
the  initial metallicity $Z$ should be above $ \sim10^{-9}$. The 0+ class should
be enriched in helium with respect to Class 0 and should exhibit low  $^{12}$C/$^{13}$C and [C/N] ratios, if observable.

Another possible  class is that of stars with none or very few Fe and $\alpha$-elements (with atomic mass number A > 24) that would exhibit the pure products of partial He-burning, \emph{\emph{i.e.}} a mixture of He, C, and O, like what the WC stars show. These stars would contain no hydrogen.
We call it composition Class "1". We do not know whether the extremely low-metallicity populations contain such stars.
The  extreme case we know at present is that of I Zw 18 with a metallicity $Z \approx 3 \cdot 10^{-4}$ \citep{Lebouteiller2013}, where numerous  Wolf-Rayet stars seem to be present   \citep{Crowther2010}. Formally, we could also consider a Class "1+", where the products of He-burning correspond to an advanced stage of He-burning. It would be characterized by a low-to-moderate amount 
of C, with a large content of O, together with some $^{20}$Ne 
 (and maybe tiny amounts of $^{24}$ Mg) produced by $(\alpha, \gamma)$ captures on $^{16}$O
and a very small amount of  $^{22}$Ne resulting from the N--destruction. This sort of object  would be like the rare WO stars, which are often observed in relatively low $Z$ regions.

\begin{figure}[t]
\begin{center}
\includegraphics[width=7cm, height=5.4cm]{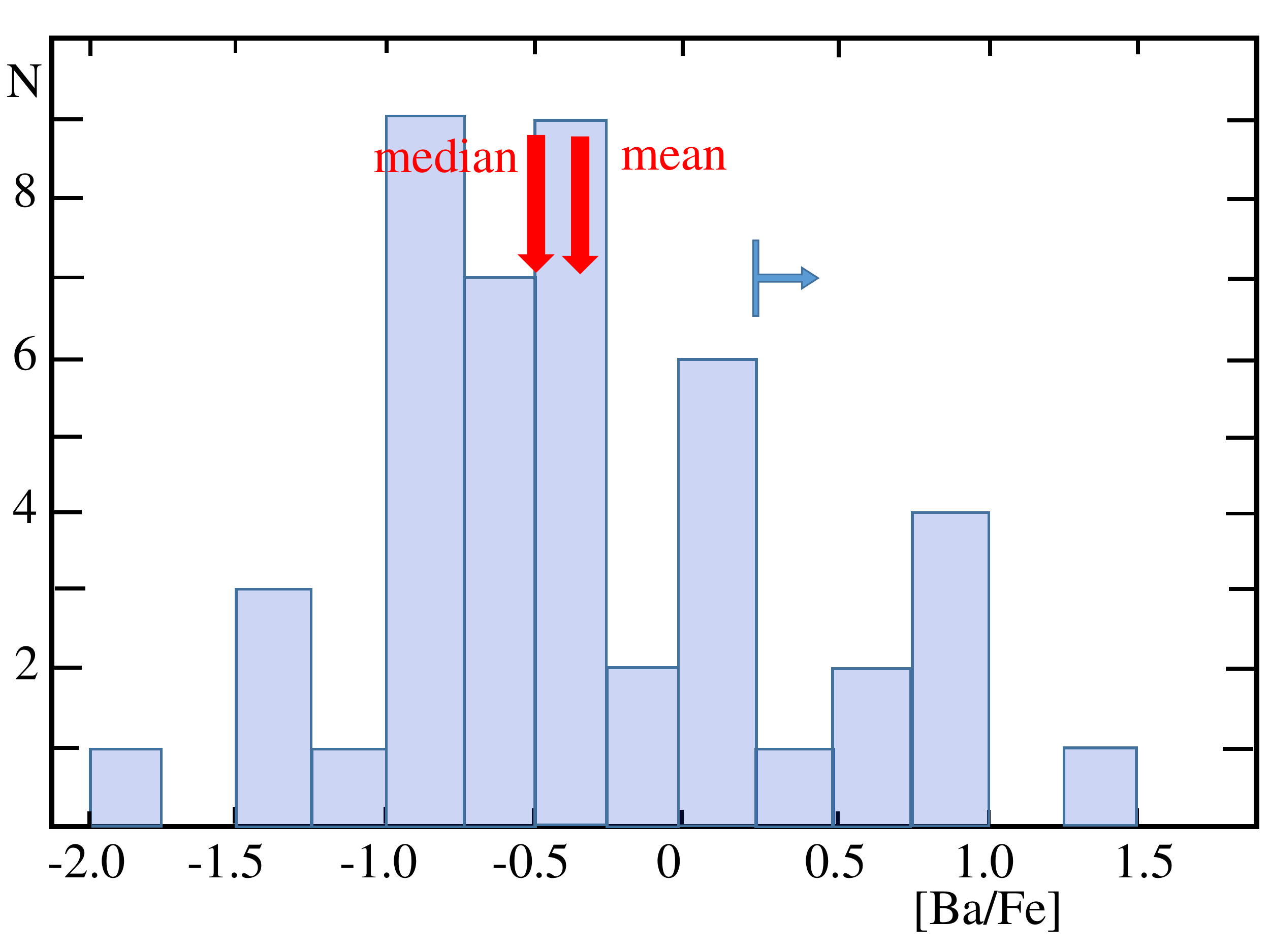}
\caption{Distribution of the ratio [Ba/Fe] for the CEMP-no stars in Table 2. Same remarks as for Fig. \ref{SR}.}
\label{BA}
\end{center}
\end{figure}

\begin{table*}[t!]  \label{table1} 
\vspace*{0mm}
 \caption{Attempt at a classification of the CEMP-no stars based on the effects of successive mixing phases
 between the H-- and He--burning 
zones and then mass loss from the source stars.  Enhancements in helium should  also be present in CEMP stars. } \label{alphalm}
\begin{center}  \footnotesize
\begin{tabular}{lcccccccccc}
Physics &Class&[Ba/Fe]&[Sr/Fe]& [Na/Fe]& [Mg/Fe] & [Al/Fe] & $^{12}$C/$^{13}$C &  [C/N] & [O/N] &  [$\frac{C+N+O}{Fe}$] \\
\hline
&  &  & \\
 & \multicolumn {10} {c}{\bf{EMP stars with none or very few heavy elements}} \\
&  &  & \\
Initial abund.            &{\bf{ 0 }}   &   \multicolumn {9} {c} {Very low abundances of all heavy elements} \\    
Prod. of H-burn.&{\bf{ 0+}} &       \multicolumn {9} {c} {He-, light  $^{13}$C-, and $^{14}$N-enrichments possible if metallicity $Z > 10^{-9}$ } \\ 
&   &   & \\   
\hline
& & &  \\
 & \multicolumn {10} {c}{\bf{CEMP stars with products of He-burning, without H (objects like WC and WO stars)}} \\
&  &  & \\
 He-burn.(WC star) &{\bf{ 1 }} &  low &     low    &   low   &    low    &  low    &   $\infty$         & $\infty$ & $\infty$    &     very high         \\ 
 He-burn.(WO star) &{\bf{1+}}&      \multicolumn {9} {c} {Same, with less $^{12}$C, more $^{16}$O,  $^{20,22}$Ne present, maybe some $^{24}$Mg} \\
 \multicolumn {10} {l} {\scriptsize \it  Note: the existence of such stars is uncertain at extremely low $Z$,  but products of 1 and 1+    intervene in mixing. } \\
&  &  &  \\
\hline 
&  &  & \\
 & \multicolumn {10} {c}{\bf{CEMP-no stars with  no s-elements and generally no  Mg and Al}} \\
&  &  & \\
 Mild $2^{nd}$ H-burn.  &{\bf{ 2 }} & $\sim$0  &   $\sim$0      &   low   &    low    &  $\sim$0   &      low-mid          & $\geq 0$ & $\geq 0 $    &     low-mid          \\
Strong $2^{nd}$ H-burn.&{\bf{2+}}& low  &       low &   low    &    low &   low    &      low                & $\leq 0.2$    &   $\leq 0.3$  &    low-mid          \\
\multicolumn{10}  {l} {\scriptsize \it  Note: Ne-Na cycle may produce Na from Ne, especially for 2+, thus leading to Class 2+Na. Mg-Al cycle unlikely.} 
 \\
&  &  &   \\
 \hline  
&  &  &  \\
 & \multicolumn {10} {c}{\bf{CEMP stars with  s-elements possible, (CEMP-no, low-s, CEMP-s)}} \\
&  &  & \\
Mild $2^{nd}$He-burn. &{\bf{ 3}}   &  low-mid &     low-mid   &    low   &    low- mid    &   low    &      high      & mid-high&mid-high &     mid-high         \\
Strong $2^{nd}$He-burn. &{\bf{ 3+}}&  mid & high &    any   &   mid-high &   any  &      high       &  $ >0.3$   &  $> 0.4$ &     mid-high       \\
More n-emission            &{\bf{3++}}  & high & high &    any     &   high     &   any   &      high   &  $> 0.3$   &  $ > 0.4$ &  high             \\
\multicolumn{10}  {l} {\scriptsize \it  Note: If Na produced in stages 2, Stages 3Na, 3+Na, 3++Na may result.  Mg present, but Al unlikely. } 
 \\
&  &  &  \\
   \hline
&  &  & \\
 & \multicolumn {10} {c}{\bf{CEMP stars with  s-elements possible, (CEMP-no, low-s, CEMP-s)}} \\
&  &  & \\
H-burn. of Class 3      &{\bf{4}}    & mid  &       mid  &    high  &     mid    &   any    &      low-mid          &  mid    &   mid     &    mid-high      \\
H-burn. of Class 3+  &{\bf{4+}}  & mid &    high &    high   & mid-high  &   mid    &      low        &  $\leq 0.3$   &   $\leq 0.4$   &    mid-high        \\
H-burn. of Class 3++ &{\bf{4++}}    & high &  high &  high   &     high   &   high  &     low       &  low   &  low  &    high                   \\
\multicolumn{10}  {l} {\scriptsize \it Note: Third passage in the H--burning zone. Na always high. Burning intensity is  increasing from 4 to 4++.} \\
\multicolumn{10}  {l} {\scriptsize \it  \hspace{5mm} Further mild mixing also leads to 4++, see text.} \\
&  &  & \\
   \hline
\normalsize
\end{tabular}
\end{center}
\end{table*}

\subsection{Classes 2}

These classes are formed by source stars in which products of He-burning (mainly C and O) have been partially mixed and processed  by the CNO cycles
in the H-burning shell, creating significant amounts of $^{14}$N. The CNO elements  transported by mixing  to the surface of the source stars
may be lost by stellar winds further, for example in the red supergiant stage as shown by \citet{Meynet2006}  or by instabilities just before the supernova explosion \citep{Moriya2015}. Some CEMP-no stars  are likely formed from a medium significantly enriched  in such ways.  This corresponds to Classes 2 and 2+, the first one having very mild H-burning, while the second shows more intense H-burning. As indicated in Table 1, such objects are expected to have no or very small amounts of s-elements like Sr or Ba. The elements
Na and Al produced by the Ne-Na and Mg--Al cycles \citep{maederlivre09} should also be generally very low or absent.  
 In Class 2, [C/N] and [O/N] still  have values above some level, say about 0, as a result of mild CNO processing. At the same time, the ratio $^{12}$C/$^{13}$C is low to medium, say $\leq$ ~10 (see, for example, Fig. 1 in Paper I).  Strong CNO processing  (Class  2+) produces
very low $^{12}$C/$^{13}$C  ratios, as well as 
 [C/N] and  [O/N] ratios below the means, which are 0.20 and  0.27, respectively. 

If  $^{20}$Ne has been produced by core He--burning in sufficiently hot objects, mixing may bring some amount of  it  into the H-burning region, allowing the Ne--Na cycle to operate 
and thus to produce some $^{23}$Na. This situation is more likely  in the case of strong nuclear processing, this would lead to a Class 2+Na. The creation of some $^{24}$Mg, if any at all, is very low in the  He-burning phase, thus  strong signatures of the subsequent Mg-Al cycles are generally not expected  in Classes 2 and 2+. However, some limited operations of the Ne--Na and Mg--Al cycles may contribute to the scatter of the values as illustrated by the  Figs. 3 to 5. Finally, we note that  the [(C+N+O)/Fe]  ratios should lie  on the low-to-medium side of the distribution illustrated in Fig. \ref{CNOSUM}, \emph{\emph{i.e.,}} smaller than about  2.5.

\subsection{Classes 3            }

The products resulting from Classes 2 and 2+ may again be mixed into the He-burning core (or in the He-burning shell in cases of advanced stages).
This leads to the synthesis of new elements  and makes  Classes 3, 3+, and 3++. Successive  $\alpha$-captures by  $^{14}$N produce  $^{22}$Ne.  Further reactions  may also occur: $^{22}$Ne($\alpha$,n)$^{25}$Mg,  and  $^{22}$Ne($\alpha$,$\gamma$)$^{26}$Mg.
The free neutrons emitted by the first one may lead to  producing some s-elements of the first peak, such as Sr and Y
\citep{Frischknecht2012}. For that it is necessary for some heavy elements of the Fe peak to be present. This is the case, since all CEMP-no stars exhibit iron, except perhaps SMSS 0313-6708, where only an upper limit is given by \citet{Keller2014}. 
In case of intense neutron emission or when the number of free neutrons per seed is high, some s-elements of the second peak may also be synthesized, for instance Ba and La. In any case, even if weak,  the reactions  may contribute to the 
scatter of the abundances (Figs. 1 and 2).
Simultaneously, the successive  $\alpha$ captures  operate and lead to the beginning of the sequence of  $\alpha$-elements:
 $^{12}$C $\rightarrow$   $^{16}$O $\rightarrow$  $^{20}$Ne. 

The presence of  nitrogen and especially of $^{13}$C in all CEMP-no stars implies that the mixing is always partial, otherwise in case of full mixing these elements would be destroyed.  We define Class 3 (cf. Table 1) as the case of a mild amount of processing into  the He-burning zone, creating more $^{12}$C
, $^{16}$O, and $^{14}$N,  leading to $^{22}$Ne
and few other $\alpha$-captures. Thus, objects in this class show none or few  s-elements. The  He-burning favors mid-to-high
[C/N], [O/N]  ratios and  a relatively  high   ($>$10) $^{12}$C/$^{13}$C ratio (within the range of CEMP-no stars). At the same time, the [(C+N+O)/Fe] ratios  are on the average slightly higher than in Class 2, mostly between 2 and 3.

\begin{figure}
\begin{center}
\includegraphics[width=7cm, height=5.4cm]{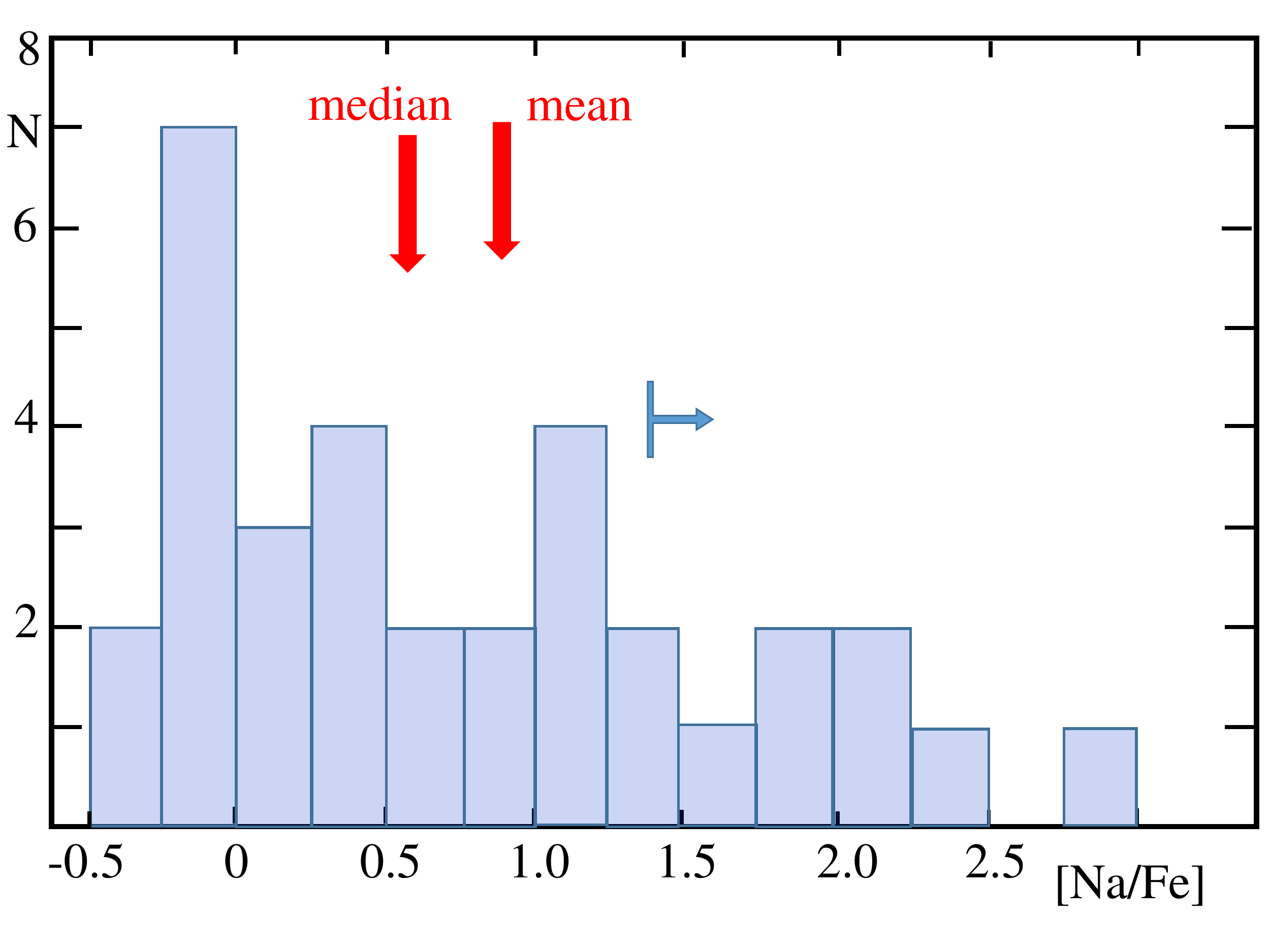}
\caption{Distribution of the ratio [Na/Fe] for the CEMP--no stars in Table 2. Same remarks as for Fig. \ref{SR}.}
\label{NA}
\end{center}
\end{figure}

We define stars of Class 3+ as those stars where the process of He-burning looks  more advanced than in Case 3. The criteria are indicated in Table 1.
The reaction producing $^{25}$Mg   leads to some significant amount of s-elements of the first peak,
typically strontium Sr. In this context,  we recall that for some time, the occurrence of Sr in very low-metallicity stars was challenging  theoretical understanding
\citep{Frebel2005}.
Figure \ref{SR} shows the present distribution of the ratios [Sr/Fe] in the CEMP-no stars. This distribution shows a rather centered peak, nevertheless with some asymmetry  due to a long tail on the right.  For the moment, we consider a star in Class 3+ if its value [Sr/Fe] is  $\geq  0.0$. At the same time, the star should present some medium-to-high abundance of  magnesium.
In Class 3++, the He-burning is even more advanced than above, and the emission of neutrons is enough to produce elements of the 
second s peak, typically barium Ba. Figure \ref{BA} shows the distribution of the [Ba/Fe] ratios for the stars of Table 2, which is rather similar to that of 
[Sr/Fe].   A value [Ba/Fe] $\geq \sim 0.25$ may correspond to Class 3++.

If Na is present in the stages corresponding to Classes 2 (being produced by the Ne--Na cycle operating from the Ne possibly 
synthesized in the core He-burning), it should then also be there in the corresponding classes  3Na, 3+Na, and 3++Na.
As we have seen, some Mg is unavoidably present in Classes 3+ and 3++, as a result of $\alpha$ captures on $^{22}$Ne. However the occurrence of Al is unlikely since it is not created in the He-burning stage.

\subsection{Classes 4}
If the elements that have experienced the previous back and forth motions again enter the H--burning region, they may produce 
 Classes 4, 4+, and 4++.  These three classes result  from further CNO-burning of the elements corresponding to Classes 3, 3+, and 3++, respectively (with or without Na). In Class 4 there are only small amounts of s-elements, CNO-burning reduces the  $^{12}$C/$^{13}$C,
[C/N], and [O/N] ratios, while the sum [(C+N+O)/Fe] remains unchanged. 
All classes of Group 4 may have high [Na/Fe] as a result of the Ne--Na cycle operating  on the Ne isotopes created in the previous He-burning stages. Figure \ref{NA} illustrates the  distribution of the [Na/Fe] ratios. This  distribution is rather flat and broad,  which may  result from the wide range of conditions leading to the Na synthesis.  The Mg content is not modified with respect to Class 3. The Mg--Al cycle may be active or not, so that any abundance of Al is possible. (The possible operation of the Mg--Al cycle does not significantly modifies the Mg-content.)

Class 4+ results from further H-burning of a medium of composition corresponding to Class 3+. Thus,  Sr shows significant excess, while this is not the case for Ba.
We have the usual effect of CNO-burning: low  $^{12}$C/$^{13}$C, [C/N], and [O/N] ratios. Na is high as before. We expect some  
moderate production of Al from the Mg--Al cycle. 

\begin{figure}[t]
\begin{center}
\includegraphics[width=7cm, height=5.4cm]{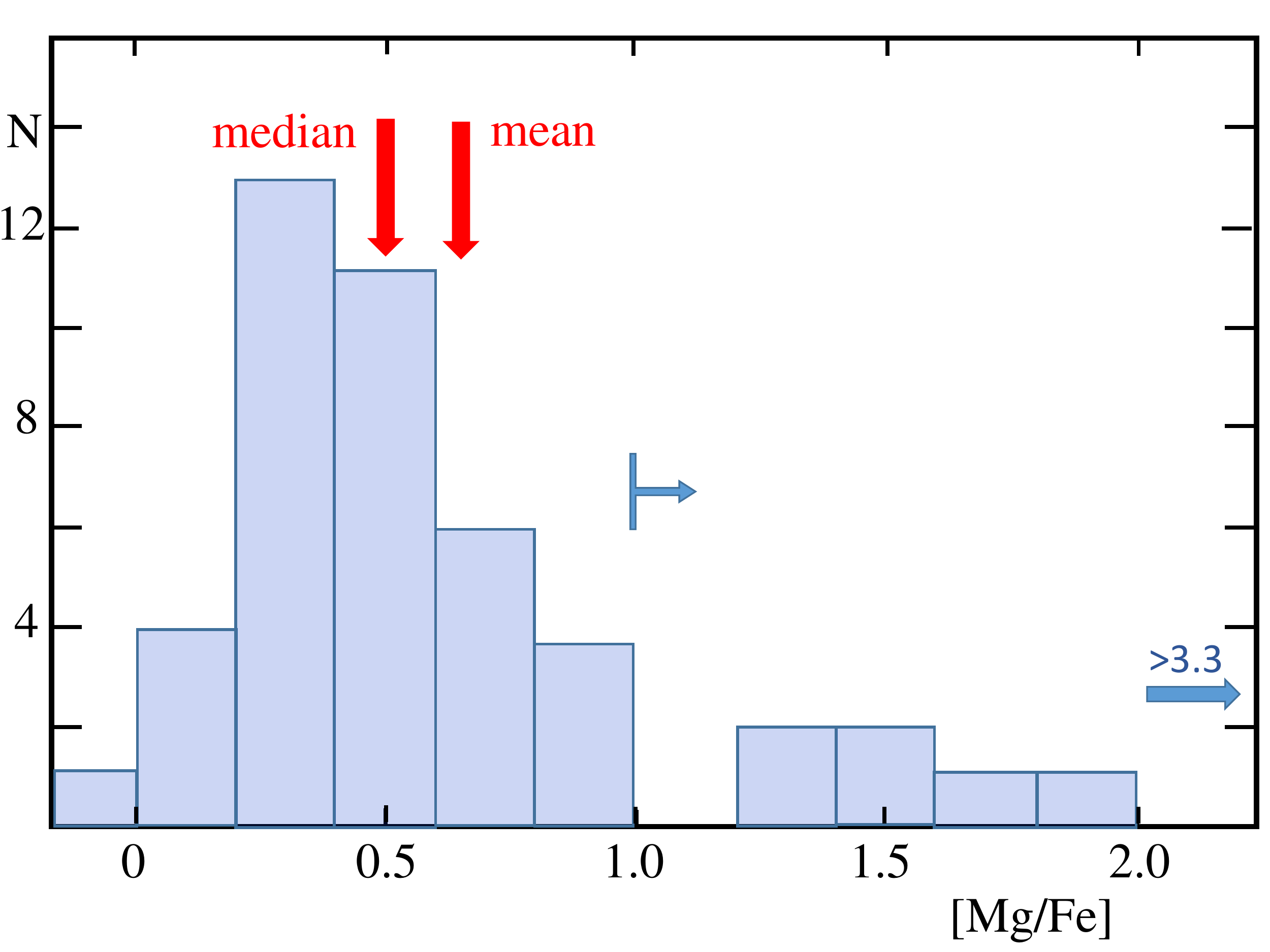}
\caption{Distribution of the ratio [Mg/Fe] for the CEMP-no stars in Table 2. Same remarks as for Fig. \ref{SR}.
The blue arrow on the right  indicates  the star SMSS0313-6708 with a ratio [Mg/Fe] > 3.3.}
\label{MG}
\end{center}
\end{figure}

\begin{figure}[t]
\begin{center}
\includegraphics[width=7cm, height=5.4cm]{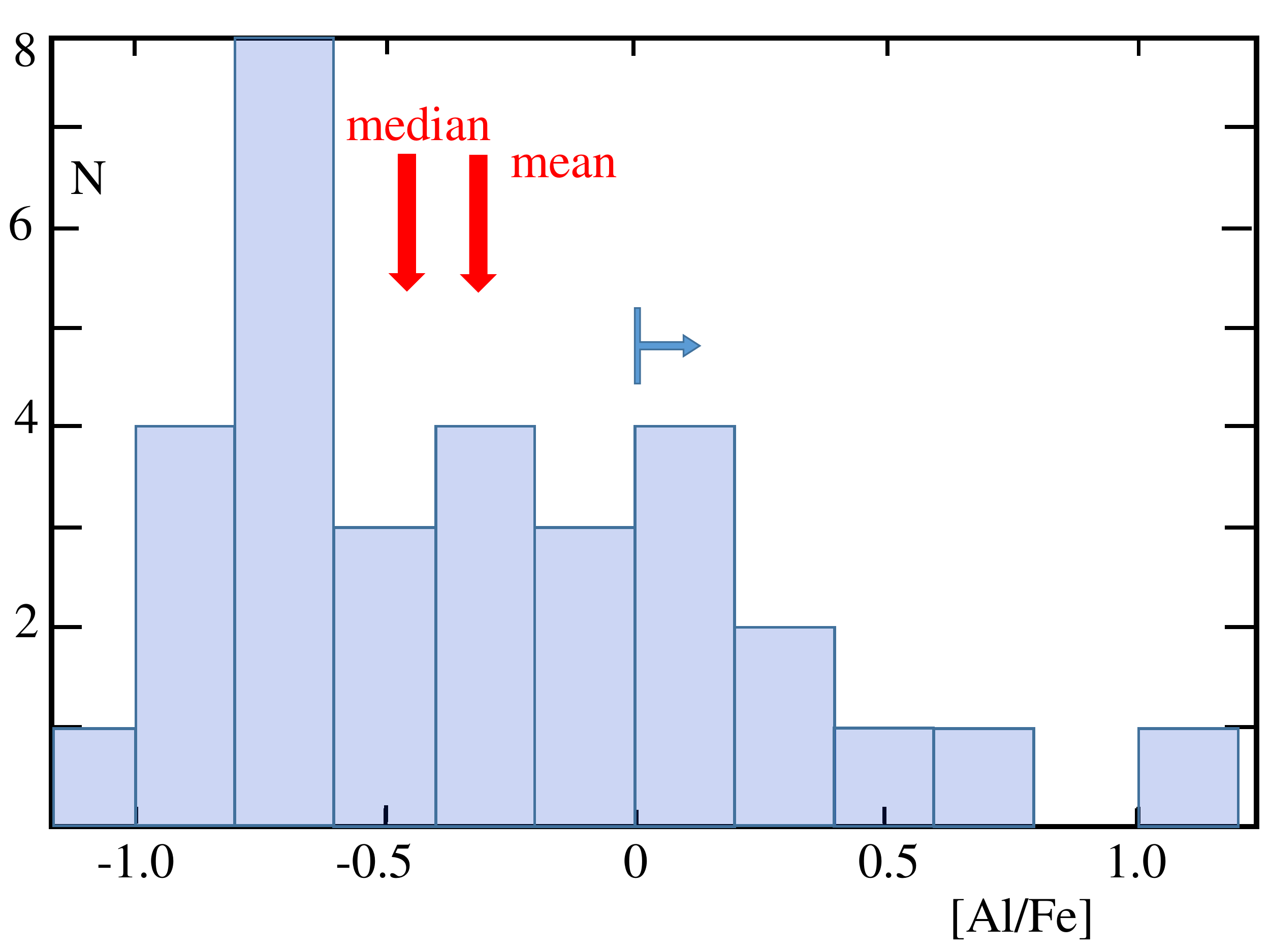}
\caption{Distribution of the ratio [Al/Fe] for the CEMP--no stars in Table 2. Same remarks as for Fig. \ref{SR}.}
\label{AL}
\end{center}
\end{figure}

Class 4++ presents both high abundances of Sr and Ba, as a result of strong
production of s-elements from intense $(\alpha,$n) captures on $^{22}$Ne. In addition, this class shows the above -mentioned effects of CNO--burning, as well as of the full operation of the Ne--Na and Mg--Al cycles, therefore the [Na/Fe], [Mg/Fe], and [Al/Fe]  ratios are all high. In this respect, we note that the 
distributions of the [Mg/Fe] and [Al/Fe] ratios are  similar (Figs. \ref{MG} and \ref{AL}), which is consistent with the fact that these elements have a  mother-daughter relation.
The distribution of the [Al/Fe] ratios is different from that of [Na/Fe],  which might be because the Al synthesis  results from a narrower 
range of conditions than for Na, since the typical temperature of the Mg--Al cycle is higher than for the Ne--Na cycle.  Moreover, the 
Mg--Al cycle requires the previous synthesis of Mg, which also demands higher temperatures than for the Ne synthesis, which is necessary for the operation of the  Ne--Na cycle.

If mixing between the H-- and He--burning regions occurs much more, the composition of the mixtures will also tend toward that of Class 4++ with
the possibility of a variety of [C/N], [O/N], and  $^{12}$C/$^{13}$C ratios. These ratios would be more like those of  Class 3++ if   He-burning was the last nuclear-burning experienced,
or more like Class 4++ if the mixture  has just gone through H-burning. This last possibility is much more likely, since to be observed, matter  has
to escape from the source star and thus has to originate in the outer layers.
As a matter of fact, we have not identified
any  3++ objects, while several stars of Class  4++ are present in the sample without any ambiguity.

The CEMP stars with s-elements are generally classified as a group different from the CEMP-no stars.
A consequence of the scheme proposed here is that some CEMP-s stars may be genetically connected to the CEMP-no stars, being
formed in a sequence of similar processes and from stars with high initial masses. Thus, further analysis of the samples of CEMP-s
stars may be useful in future  to extract  the objects that in fact belong to the genetic group of the so-called CEMP-no stars, likely  stemming from massive stars in view of the short timescales associated to objects with [Fe/H] lower than -3.0 \citep{Chiappini2008,Matteucci2012}.

\subsection{A  word of caution and specific remarks}

The rich composition variety of CEMP-no stars results, as we are suggesting, from  mild mixing between 
the H- and He-burning regions of massive stars with mass loss, which allow the new elements to escape into the ambient medium. The range of possible initial stellar masses is wide, say all masses above about 10 M$_{\odot}$. Thus, the reduction of the many complex processes occurring in stars of different masses, rotation velocities, mass loss,  and metallicities to a limited number of simple classes is a kind of wager. We examine whether such a  complex reality encounters any difficulty when being forced into such a simple scheme. As a matter of fact, problems rarely occur.
We would also like to point out that the values of the limits between the various classes may change a bit when larger samples of CEMP-no stars become available and when observations become more accurate. Although mainly qualitative at present, the classification
is nevertheless very constraining, owing to 
the number of criteria to be satisfied  for each class. In future, it may become more quantitative, if progress in stellar models and observations allows it.

There is a fundamental difference between elements like C, N, O, Ne, Na, Mg, and Al and the s-elements. The first ones in CEMP-no
stars are primary elements,  (\emph{{{\emph{i.e.,}}}} made from the initial H and He), while the second ones are secondary elements. To be created, the s-elements require seed elements of the Fe peak, by which free neutrons are captured.   The behavior of these two kinds of elements is different in  models of  galactic chemical evolution.

 Most interestingly,  \citet{Hansen2015} have found a "floor" -- {{\emph{\emph{i.e.},}}} a plateau independent of [Fe/H]  (for [Fe/H] $< \sim  -3.0$) -- in the absolute Ba abundances of CEMP stars $A(Ba)=\log 
\epsilon(Ba)+12.0 \approx -2.0$, where $\epsilon$ indicates the abundances in number.  This result  implies that for the lower [Fe/H] ratios, the observed
values of [Ba/Fe] become increasingly higher. Here, we only have CEMP-no stars, i.e., those  with no overabundance of Ba according to the formal definition.
Nevertheless, in the limited range of [Ba/Fe] ratios concerned (normally [Ba/Fe]<1), the data of Table 2 support this result. 
There are  eight stars that show an excess of s-elements with [Sr/Fe]>0  (Fig. 1): BS 16929-005, 
CS 22949-037, HE 0100-1622, HE 0233-0343, HE 1300-2201, HE 1327-2326, HE 1330-0354, and 53327-2044-515. They all have negative
values of [Ba/Fe], except the two largely most metal-deficient ones, HE 0233-0343 and HE 1327-2326, where the upper limits are highly positive.

How do we interpret the Ba floor found by \citet{Hansen2015}? We suggest the following.  At the lowest [Fe/H]  values, the number of seeds per free neutron becomes small enough that the neutron flux is not so much reduced once the elements of the first peak of the s-process are synthesized.
 Thus,  further  n-captures may lead to the second neutron peak (Ba).

We have considered  that the  s-elements are produced by $(\alpha,$n) captures on $^{22}$Ne, 
as is clearly suggested by the nucleosynthertic models
of massive stars at very low metallicities by \citet{Frischknecht2012}.  We may wonder whether the destruction of $^{13}$C by the $^{13}$C($\alpha$,n)$^{16}$O reaction
in the stages corresponding to Classes 3 may also produce s-elements. If this happens, Classes   3 and 4,  which normally 
contain no s elements (cf. Table 1),  could also present some of them.
However, this channel of s--element production should be small,  because $^{14}$N is always present at the same time as $^{13}$C, and it acts like a poison by capturing the free neutrons, thus preventing the production of many s-elements.  Nevertheless, some traces of s-elements are possible in all products of  He--burning, and this may also be part of why
the distributions of [Sr/Fe], and [Ba/Fe] values are rather broad (Figs. \ref{SR}, \ref{BA}).

Finally, we comment on the relations between the Mg content and the s--elements. The synthesis of $^{25}$Mg implies the synthesis of s-elements (at the same time $^{26}$Mg is also  created). However, at very low $Z,$ some small synthesis of $^{24}$Mg is also possible,
see for example Table 4 by \citet{Meynet2006}. Thus, depending on the physical conditions (mass, $Z$, rotation, mass loss, etc.),  some small amount
of Mg could be present, without  there being any s-elements.
Is the opposite possible with some s-elements present without a significant Mg abundance? This possibility is very marginal, it could  occur  in the case of a small production of s-elements by $\alpha$-captures on  $^{13}$C.

\section{Proposed classification of CEMP-no stars}

Table 2 summarizes the useful  abundance ratios for a classification of  the  46 CEMP-no  or related stars  from Paper I and \citet{Hansen2015}.
The proposed classes are given in Column 5. 
 There are clearly not enough data for 16 objects, to which a class cannot be reasonably assigned.  A classification appears  possible for 30 stars.
In several cases and despite some missing data, we have attempted to make the classification. Four cases 
  noted by ":"  are very uncertain  because of generally insufficient data.  We start by  the less advanced evolutionary stages.

\begin{table*}[t!]  
\vspace*{0mm}
 \caption{Abundance data for the sample stars considered in the attempt to assign a class } \label{alphalm}
\begin{center}\scriptsize
\begin{tabular}{ccccrrrrrrrrrrr}
Star  &  $ T_{\mathrm{eff}}$ & $ \log g $ &  [Fe/H] & {\bf{Class}} & $^{12}$C/$^{13}$C  & [C/N] & [O/N] & [$\frac{C+N+O}{Fe}$] & [Na/Fe] & [Mg/Fe] &[Al/Fe]   & [Sr/Fe]  &   [Ba/Fe]   & ref\\
&   &  & & &   &  & \\
\hline
&   &  & & &   &  & \\
BD +44 493             & 5510  & 3.70  & -3.68      & {\bf{3}}        & >30                 & 0.99          &     1.27    & 1.48           & 0.27   & 0.52         & -0.57      & -0.26        &   -0.59           & 1, 5\\
BS 16929-005      & 5229   & 2.61  & -3.34      & {\bf{3}}+      &$ >$7         & 0.67         &  -               &  -                & 0.03    & 0.30        &   -0.72     & 0.54         &   -0.41           & 1\\
CS 22166-016       &  5250  & 2.0    &   -2.40    & -                   & -                & -                 &  -               & -                 & 0.37    & 0.68        &-                & -1.01       &   -0.37            & 2\\
CS 22877-001       &  5100 &   2.2 &  -2.72       & {\bf{3}}        & $>$10      & 1.00          & -                &  -                & -0.24   & 0.29        & -0.72      & -0.12         &   -0.49           & 2,3 \\
CS 22878-027       & 6319 & 4.41  & -2.51       & -                    &-                 & $>$-0.20 & -                &      -             & -0.17   &  -0.11     & -              & -0.26        &  $<$-0.75       & 1\\
CS 22885-096      &   5050 & 1.9  &  -3.66        & -                   & -                & -                & -                & -                  & -           & 0.52        & -0.78      & -1.59        &   -1.44            & 2\\
CS 22945-017      &   6400 & 3.80  &  -2.52     & {\bf{4++}}    & 6               & 0.04          & $<$0.12  & $<$2.33     & -          & 0.61        &-                & -                &  0.55             & 2b-,3\\
CS 22949-037      & 4958 & 1.84  & -3.97        & {\bf{4+}}      & 4                & -1.10         &  -0.18      & 1.86          & 2.10     & 1.38        & 0.02         & 0.55         &   -0.52          & 1\\
CS 22956-028      &   6700 & 3.50  &  -2.33     & -                   & 5                & -0.01         & $<$0.62  &$<$2.31    & -           & 0.58        &-                & -                &   0.16           & 2b-,3\\
CS 22957-027      & 5170 & 2.45  & -3.19        & -                    & 6               & 0.52         &  -                & -                 & -           & 0.30        & -0.10       & -0.86        &   -0.80          & 1\\
CS 22958-042      & 6250 & 3.5 & -2.85           & {\bf{4}}         & 9                & 1.00         & -0.80         & 2.68          & 2.85     & 0.32       &-0.85        & -0.20         & $<$-0.53     & 2, 3 \\
CS 29498-043       & 4639 & 1.00  & -3.49       &{\bf{4}}          &  6               & -0.40       &   0.13         &  2.31        & 1.47     & 1.52       & 0.34         & -0.35         & -0.45            & 1\\
CS 29502-092        & 5074 & 2.21  & -2.99     & {\bf{3}}          & 20             & 0.15        &  -0.06          &  0.83        & -            & 0.28      & -0.68         & -0.15         &   -1.20         & 1, 2\\
CS 30314-067        &  4400 & 0.7  &  -2.85      & -                     & -                & -0.70      & -                 & -                 & -0.08     & 0.42      &-0.10         & -0.37           &   -0.57        & 2\\
CS 30322-023         &   4100 & -0.30  &  -3.39 & {\bf{4++}}     & 4               & -2.11      & -2.28         & 1.84          & 1.04       & 0.80      &-                 & -                   &   0.59         & 2b-,3\\
CS 31080-095         &   6050 & 4.5  &  -2.85     &{\bf{ 3,3+}}    & $>$40     & 1.99       & 1.65           & 2.47          & -0.28      & 0.65      &-0.95          & -0.29          &   0.05         & 2b-,3\\
G 77-61                     &   4000 & 5.05  &  -4.03   & -                    & 5              & 0.00       & -                 & -                  & 0.60       & 0.49       & -                &$ <$0           &   $<$1       & 2, 3\\
HE 0007-1832         &   6515 & 3.8  &  -2.72     & -                    & -               & 0.78       & -                 & -                  & -              & 0.79       &  -               &  -                   &0.23           & 3\\
HE 0057-5959        &  5257 & 2.65 & -4.08      & {\bf{2+Na}} &$>$ 2        & -1.29     &  $<$0.62  &   <2.56        & 1.98       & 0.51       & -                &  -1.06         & -0.46          & 1\\  
HE 0100-1622            & 5400  & 3.0  & -2.93      & {\bf{3+}}        &  13           & 0.85          &   <0.40    &    <2.49        & >1.00  & 0.64        & 0.46      & 0.25        & <-1.80           & 6\\
HE 0107-5240        &  5100 & 2.20 & -5.54      &{\bf{3}}          & $>$50     & 1.42       &  -0.13       &   3.38           & 1.11       & 0.26       & $<$-0.26 &  $<$-0.52  & $<$0.82     & 1, 2\\
HE 0134-1519             & 5500  & 3.2  & -3.98      & {\bf{3}}        & >4                 & >0          &     -    &   <2.68         & -0.24   & 0.25         & -0.38      & -0.30        &  <-0.50           & 6\\
HE 0146-1548       & 4636 & 0.99  & -3.46       & -                    & 4              & -              &   -              &  -                  & 1.17        & 0.87      & 0.14          & -0.38          &  -0.71         & 1\\
HE 0233-0343             & 6100  & 3.4  & -4.68      & {\bf{3+}}        & >5                 & >0.68          &     -    &  <3.85         &< 0.50   & 0.59         & <0.03 &    0.32        &   <0.80           & 6\\
HE 0440-1049             & 5800  & 3.5  & -3.02      & {\bf{3}}        & -                 & >0.07          &    -    & <2.46    &-0.04          & 0.79         & -0.57      & -0.30        &   -1.27           & 6\\
HE 0557-4840       & 4900 & 2.20  & -4.81       &{\bf{3}}          &  -              & $>$0.70 &  $>$1.30  & $<$2.13   & -0.18       & 0.17       & -0.65         &  $<$-1.07  &   $<$0.03  & 1\\
HE 1012-1540        & 5745 & 3.45  & -3.47       &{\bf{3Na}} & -               & 0.97        &  1.00         & 2.21           & 1.93        & 1.85       & 0.65         & -0.37           &   -0.25       & 1\\
HE 1150-0428        & 5208 & 2.54  & -3.47       & -                    & 4             & -0.15       &  -                &  -                & -                & 0.41       &  -              &  -0.12          &   -0.48       & 1\\
HE 1201-1512        & 5725 & 4.67  & -3.89       &{\bf{3}}           & $>$20   & $>$0.11  &  -               & $<$2.43    &-0.33        &  0.24      & -0.73       & $<$-.87       &  $ <$0.05   & 1\\
HE 1300+0157       & 5529 & 3.25  & -3.75       & {\bf{2}}          & $>$3       & $>$0.60 &  $>$1.05 & $<$1.62    & -0.02       & 0.33       & -0.64       & -1.36            &   -0.85       & 1, 2\\
HE 1300-0641        &   5308 & 2.96  &  -3.14    & -                     & -              & -               & -                & -                 & -                & 0.04       & -1.21       & -0.59            &   -0.77      & 2\\
HE 1300-2201        &   6332 & 4.64 &  -2.61     & -                     & -              & -               & -                & -                 & -                & 0.29       &-0.92        & 0.28             &   -0.04      & 2\\
HE 1310-0536     & 5000  & 1.90  & -4.15      & {\bf{2+}}        & 3         & -0.84          &   <-0.40    &     <2.76       & 0.19   & 0.42         & -0.39      & -1.08 &   -0.50     & 6\\
HE 1327-2326        & 6180 & 3.70  & -5.76       &{\bf{4++}}      &  $>$5     & -0.30        &  -0.86       &  4.07         & 2.48          & 1.55       & 1.23        & 1.04            &   $<$1.46  & 1, 2\\
HE 1330-0354        &  6257& 4.13  &  -2.29      & -                    & -               & -                & -               & -                 & -                 & 0.32      & -0.93        & 0.01            &  -0.47       & 2\\
HE 1410+0213       &   4890 & 2.00  &  -2.52    & {\bf{2+:}}      & 3             & -0.61         & -0.38        & 2.55         & -                 & 0.33       & -               & -                   &   0.06       & 2b-,3\\
HE 1419-1324        &   4900 & 1.80  &  -3.05    &  {\bf{4++:}}   & 12           & 0.29          & $<$-0.28 &$<$1.48  & -                  & 0.53       & -               & -                  &   0.88          & 2b-,3\\
HE 1506-0113        & 5016 & 2.01  & -3.54       & {\bf{3Na}}     &  $>$20  & 0.86          &  $<$1.71  &  $<$2.13 & 1.65           & 0.89      & -0.53       & -0.85          &   -0.80         & 1\\
HE 2139-5432        & 5416 & 3.04  & -4.02       & {\bf{3Na}}     & $>$15   & 0.51          &   1.07        &  2.99        & 2.15           & 1.61      & 0.36        & -0.55          &   $<$-0.33   & 1\\
HE 2142-5656        & 4939 & 1.85  & -2.87       & -                     & -              & 0.41          &   -               &  -               & 0.81         & 0.33       & -0.62      & -0.19          &   -0.63           & 1\\
HE 2202-4831        & 5331 & 2.95  & -2.78       &-                      & -              & -                 &  -               & -                & 1.44          & 0.12       &-               & -0.85          &   -1.28          & 1\\
HE 2247-7400        & 4929 & 1.56 & -2.87        & -                     & -             & -                  &   -               &   -              & 0.82         & 0.33       & -              & $<$-0.15    &   -0.94         & 1\\
HE 2331-7155             & 4900  & 1.50  & -3.68      & {\bf{4}}        & 5                 & -1.23          &    <-0.87    &    <1.82        & 0.46   & 1.20         & -0.38      & -0.85        &   -0.90           & 6\\
Segue 1-7              & 4960 & 1.90  & -3.52        & {\bf{3}}          &$>$50    & 1.55           & $<$1.46    &  $<$2.21 & 0.53         & 0.94      & 0.23         & -1.39          &   $<$-0.96  & 1\\
SMSS 0313-6708  &   5125 & 2.3  &  <-7.1       & {\bf{4++:}}    & $>$4.5   & -                 &  -                 & >4.31      & -                 & $>$3.3 & -                & -           &   -         & 4\\
53327-2044-515  & 5703 & 4.68 & -4.05         &{\bf{ 3+:}}       & $>$2      & -                 &  -                 & -               & 0.14          & 0.40      & -0.17        & 1.09           &   $<$0.34   & 1\\
   &   &  & & &   &  & \\

   \hline
\end{tabular}
\end{center}
\vspace*{0mm}
\scriptsize{Ref: 1. \citet{Norris4}; 2. \citet{Allen2012};  3. \citet{Masseron2010}; 4. \citet{Keller2014} give values of [Li/H], [C/H], [Mg/H],  and [Ca/H]
and upper limits  in the case of  other  elements for this star. 5. \citet{Ito2013} give a lower limit for $^{13}$C/$^{12}$C.  6. \citet{Hansen2015}.}
\vspace*{0mm}
\end{table*}

\subsection{Stars in Classes 2 and 2+}
This class is characterized by products of He-burning that undergo some first mixing and processing in the H-burning shell and then escape
by mass loss. HE0057-5959   shows no sign of excess in [Ba/Fe] and [Sr/Fe], because this last ratio is  very low. [Mg/Fe]  is close to the mean, but
 there is a strong excess in the [Na/Fe] ratio.  [C/N] is low, while [O/N] is given by an upper limit. 
$^{12}$C/$^{13}$C is above 2. This star  shows the effects of H-burning, and we  assign a Class 2+Na to it.
This object (like CS 22949-037 in Class 4+) was cataloged as a NEMP-star (nitrogen enhanced metal poor) by  \citet{Johnson2007}, but it was considered as a CEMP-no star by \citet {Norris4}.
We support this  last appreciation. Although both Classes 2+ and 4+ have low [C/N] values, they also have large differences 
in  parameters, such as [Sr/Fe], [Mg/Fe], and [Al/Fe], which  consistently places these two stars in different classes.

HE 1300+0157 is a good example of an object with both very low [Sr/Fe] and [Ba/Fe] ratios.   This is consistent with the low value of the [Mg/Fe] ratio. The ratios [Na/Fe] and [Al/Fe] are well below the mean, and  there is certainly no evidence at all of the Ne-Na and Mg-Al cycles.   Both [C/N] and [O/N] have rather high lower bounds.  $^{12}$C/$^{13}$C is greater than 3, which is not very constraining.
 The available data support Class 2.

HE 1310-0536 shows  values  close or below the mean for all ratios [Sr/Fe], [Ba/Fe], [Na/Fe], [Mg/Fe], and [Al/Fe], indicating no excess of
s-elements and no evidence of the Ne--Na and Mg--Al cycles.
The ratios  [C/N],  [O/N], and $^{12}$C/$^{13}$C (which we call here "the CNO triplet")
 are all very low, particularly  $^{12}$C/$^{13}$C and [C/N]. These various abundances  are all consistent with Class 2+.

HE 1410+0213  was considered as a low s--star by \citet{Allen2012}, however the value of [Ba/Fe] is 
only slightly above the mean (Fig. \ref{BA}).   The CNO triplet  has low values, clearly indicating efficient H--burning. That [Mg/Fe] is below the mean 
 favors Class 2+, rather than one of the classes of Group 4. However, the  data on Na, Al, and Sr are missing, and it is difficult to assign a class to this star
with confidence.
We   add the sign of uncertainty. 

\subsection{Stars in classes 3, 3+, and 3++}

 In Classes 3, 3+, and 3++, we see  products that
are  again  mixed into an He--burning region, allowing new elements to  experience $\alpha$-captures and to synthesize 
 $^{16}$O, ($^{18}$O), $^{20}$Ne,  $^{22}$Ne, and possibly $^{24, 25, 26}$Mg, as well as s-elements of the first (Sr, Rb, Y, Zr, etc.) and second (Ba, Cs, La, Ce, etc.) peaks before they escape in stellar winds.

The star BD +44 493 shows low-to-medium values of all heavy elements  Na, Mg, Al, Sr and Ba relative to Fe. It has significantly  high [C/N] and [O/N] ratios. 
The rather high lower limit   $^{12}$C/$^{13}$C > 30 confirms that  Class 3 is appropriate. Radial velocities have been intensively studied, and no significant variations were found  for this star \citep{Carney2003,Norris4}.

BS 16929-005 has a mean [Ba/Fe], but a high [Sr/Fe], thus it gets a  symbol "+". The elements Na, Mg, and Al are low. [C/N] is moderately high,
and  $^{12}$C/$^{13}$C > 7. These various ratios are clearly consistent with a Class-3+ designation.

 CS 22877-001 also does not show excesses of the  s-elements, which have medium values. The other  heavy elements Na, Mg, and Al, have low-to-medium values. The rather high [C/N] and the lower bound of 10 
 for $^{12}$C/$^{13}$C  suggests Class 3, however  it 
would be useful to confirm the O abundance.

 CS 29502-092  has a very low [Ba/Fe] ratio, while [Sr/Fe] is slightly above the mean, but not with significant excess. Mg and Al are well below the mean.
 Data on Na are missing. The values of [C/N] and [O/N] are close to 0, while
$^{12}$C/$^{13}$C= 20, which is medium-to-high in the context of CEMP-no stars.  This star shows  the properties  of Class 3, but not  far from 
Class 2.

\begin{figure}[t]
\begin{center}
\includegraphics[width=9.0cm, height=7.0cm]{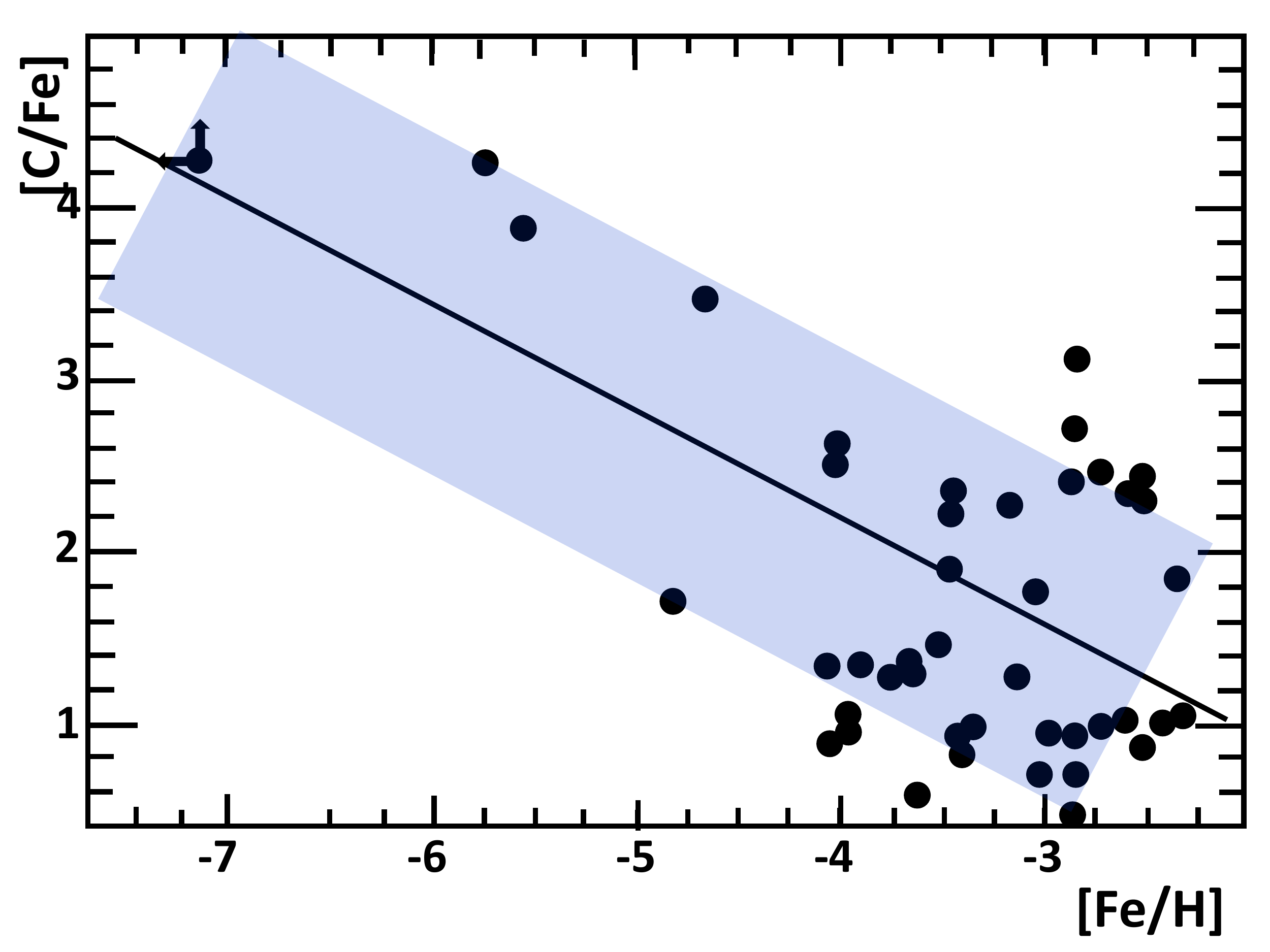}
\caption{Rather poor relation between [C/Fe] and [Fe/H] for CEMP-no stars. 
The light blue band indicates the standard deviation in [C/Fe] of  0.99 on both sides of the least square average relation.}
\label{CFE}
\end{center}
\end{figure}

CS 31080-095  has values of 
[Sr/Fe]  and [Ba/Fe] that are slightly above the mean, however not high enough to give  significant excesses (see Figs. \ref{SR} and \ref{BA}). It has  very low Na and Al content and a value of  [Mg/Fe] just in the mean. The  high CNO triplet places it in Class 3, not far from 3+ in view of the Sr and Ba abundances. That this Class 3 to 3+ implies the visibility of products of the second He-burning is consistent with the remark by \citet{Allen2012} that it is one of the few  O-rich stars  with little Na.

HE 0100-1622 shows an excess of [Sr/Fe], but not for barium, so this clearly implies a sign "+". At the same time, it exhibits [C/N] much above the mean, also with a positive [O/N]. Together with the middle-to-high $^{12}$C/$^{13}$C =13, this supports Class 3+. 
This is consistent with a middle-to-high Na, as well as with evidence of some  modest effect of the Mg--Al cycle.

For HE 0107-5240, we only have upper bounds for [Ba/Fe] and [Sr/Fe] for this very low-metallicity star. \citet{Norris4} notice that although it cannot be formally classified as a CEMP--no star, it shares all its characteristics.
 In view of the relatively low value of the  upper bound for [Sr/Fe], this star does not  get "+" or "++" symbols. Norris et al. also point out that this star has  a long binary period (P=150 yr). Na is slightly above the mean. We notice a high lower bound for  $^{12}$C/$^{13}$C, which determines a Class 3.  [(C+N+O)/Fe] is also high.
The point in the [O/N] vs. [C/N]  diagram (Paper I), although a bit low, is clearly far from CNO equilibrium, which is consistent with the presence of products of He-burning.

HE 0134-1519 presents values of [Sr/Fe] and [Ba/Fe] close to or below the mean.  [Na/Fe] is much below the mean,  and there is also no sign
of the Mg--Al cycle. The CNO triplet has only lower bounds. Nevertheless, we propose a Class 3, because this star is neither  a Class 4 in view of the low Na
nor a Class 2 in view of the average Sr and Ba contents. The Class 3 is consistent with a positive [C/N], moreover the upper bounds [N/Fe]<1 and 
[O/Fe]<2.90 may also  suggest a positive [O/N].

HE 0233-0343 clearly has an excess of [Sr/Fe], which implies a "+"sign.  For [Ba/Fe], we have an upper bound of 0.80, which allows an excess 
 and thus the 
possibility of having a "++" object. The data for Na,
Mg and Al (upper bound) indicate medium values with no clear evidence of the Ne--Na and Mg-Al cycles. The lower bound for $^{12}$C/$^{13}$C > 5
 is not very constraining. The positive  [C/N] ratio supports Class 3+ (or 3++),  which is also consistent with the two upper bounds for [N/Fe]< 2.80 and [O/Fe]<4.00, suggesting a positive  [O/N] ratio.

HE 0440-1049 is the  CEMP-no star with the lowest [C/Fe] ratio \citep{Hansen2015}. It shows a value of [Sr/Fe] close to the mean, while clearly [Ba/Fe] is very low.  [Na/Fe] is also low, [Mg/Fe] close to the mean,
while [Al/Fe] is rather low to medium. This indicates an absence of efficient Ne--Na and Mg--Al  cycles. There is  a lack of data for the CNO triplet: we only safely know that [C/N] is positive, which supports a Class 3.
This agrees with the upper bounds [N/Fe]<0.62 and [O/Fe]< 2.50, which suggest a positive [O/N] ratio.

HE 0557-4840 has a very low value of [Sr/Fe].  [Ba/Fe] is given by an upper bound. The values of [Na/Fe], [Mg/Fe], and [Al/Fe] are below the mean,  and there is clearly no evidence of the Ne--Na  and Mg--Al cycles.
[C/N] and [O/N], given by lower limits, are largely positive, indicating a dominance of the products of He-burning. $^{12}$C/$^{13}$C is missing, nevertheless the presently available   data
support a Class 3.

HE 1201-1512 shows no evidence of significant s-elements in view of the upper bounds given by \citet{Norris4}, nor of the Ne--Na and Mg--Al cycles.  $^{12}$C/$^{13}$C is 
relatively high, and we have a positive lower
bound for [C/N]. These properties allow us to  assign a Class 3 to this star. 

HE 1506-0113 also presents no
 excess of s-elements, with values well below the means. Mg is a bit above the mean, while there is clearly a high [Na/Fe] ratio. However, there
is no evidence of the Mg--Al cycle.
 $^{12}$C/$^{13}$C  and [C/N] are  high, while   [O/N] has a  high upper bound.  We place this star in Class 3Na.

 For HE 2139-5432, both Sr and Ba are in the mean or below. [Na/Fe] is clearly high, as is [Mg/Fe]. In addition to the Ne--Na cycle, there is  an evidence of the operation of the Mg--Al cycle, in view of the relatively high value of [Al/Fe].  $^{12}$C/$^{13}$C has a
lower limit  of 15, while [C/N] and [O/N] are above the means. These properties place the star in Class 3Na. That among two stars 
of Class 3Na, one shows a signature of the Mg--Al cycle and the other not may possibly be related to a difference in the  mass of the source star,
 in the sense that the signature of the Mg--Al cycle may suggest a higher mass.

Segue 1-7 shows very low [Sr/Fe] and [Ba/Fe] ratios. This star has both [Mg/Fe] and[Al/Fe]  somehow above the mean, but not with significant excesses. This  might suggest 
a weak occurrence of the Mg--Al cycle. We remark that  [Na/Fe]
is slightly below  the mean, however the differences are within about 0.3 dex.  It has  very high  $^{12}$C/$^{13}$C  and [C/N]  ratios (the upper bound given for [O/N] is also high),  thus we assign 
 a Class 3 to this star. 

The star 53327-2044-515 is considered as closely related to the CEMP-no star by \citet{Norris4}. There is a very strong excess of the ratio  [Sr/Fe].
The upper bound of  [Ba/Fe] is compatible with a slight excess. Thus, we retain the  symbol  "+", although it may not be far from  "++". 
 There is no evidence of  active Mg--Al  and Ne--Na cycles. The lack of data about [C/N] and [O/N] together with the fact that $^{12}$C/$^{13}$C
only has a lower bound prevents a reliable classification.  Provisionally, we suggest a Class 3+ with a mark of uncertainty.

\begin{figure}[t]
\begin{center}
\includegraphics[width=9cm, height=7cm]{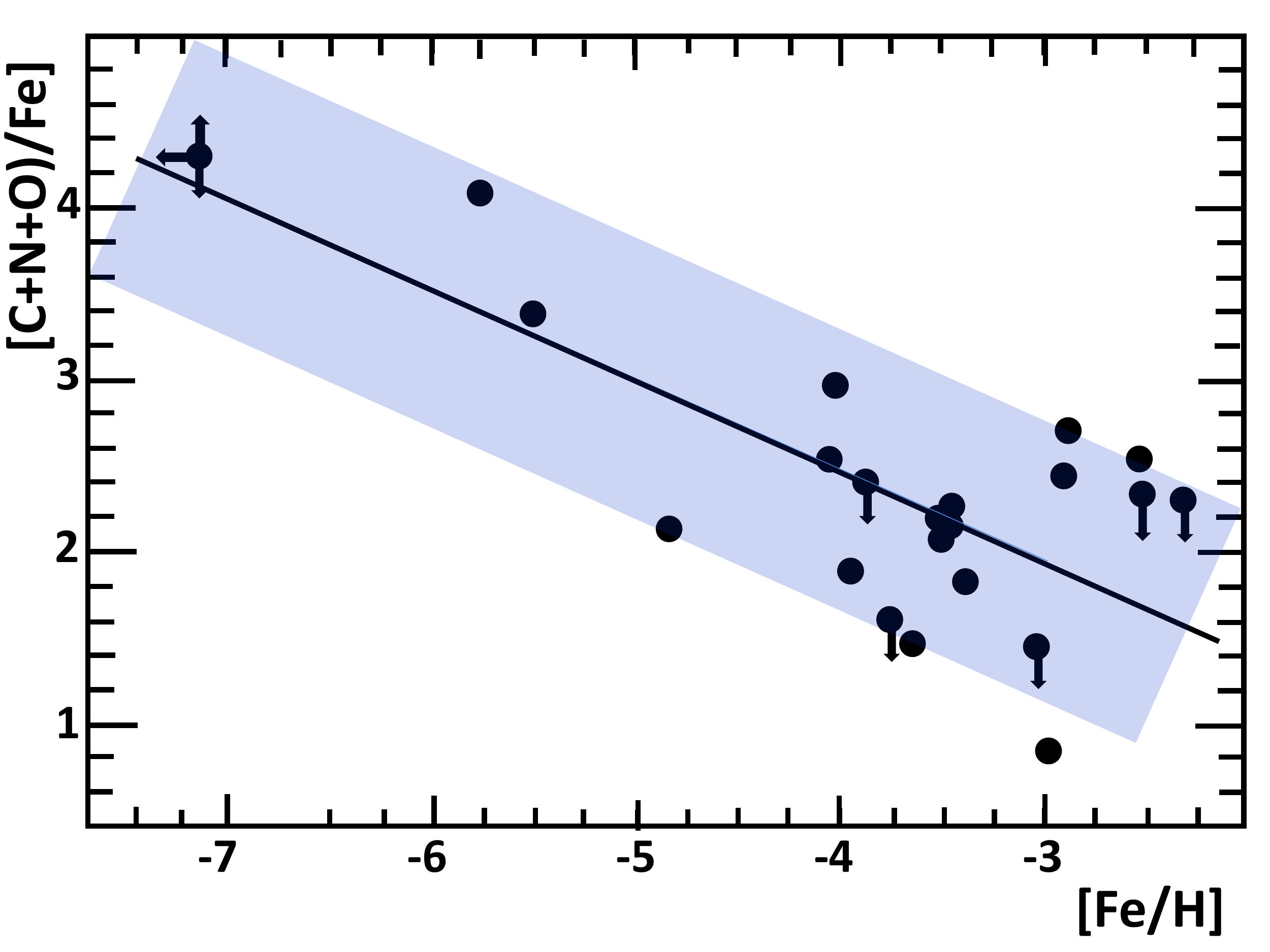}
\caption{Relation between [(C+N+O)/Fe] and [Fe/H] for CEMP-no stars. 
The light blue band indicates the standard deviation in [(C+N+O)/Fe] of  0.88 on both sides of the least square average relation.}
\label{CNOSUM}
\end{center}
\end{figure}

\subsection{The remarkable stars in Classes 4, 4+, and  4++}

These classes contains stars, where products of He-burning are mixed for the second time (or even more) into the H-burning shell.
There is no ambiguity for the definition of Classes 4+ and 4++. The differences between Classes 4 and 2Na are small (two classes of mild H-burning), it rests on the very low values of [Sr/Fe] and [Ba/Fe] in Class 2 and on the
higher value of [Mg/Fe] in Class 4.  The ratio [(C+N+O)/Fe] is generally higher in Classes 4, but this is only a statistical indication (see Fig. \ref{CNOSUM}).

CS 22945-017 was  considered to be a "b-" star by \citet{Allen2012}, which means with [Ba/Fe] between 0 and 1 and no europium (r--element). This star shows a relatively high [Ba/Fe] ratio, and at the same time
relatively low values of the CNO triplet. Thus, we may consistently put  it in 
Class  4++,  nevertheless indications about Sr, Na, and Al would be desirable for a confirmation. 

CS 22949-037 shows a low-to-medium value of [Ba/Fe], but a high value of [Sr/Fe]  with   high  [Na/Fe]  and [Mg/Fe] ratios. It exhibits low values of  the CNO triplet, so that it was sometimes cataloged as a NEMP-star  by \citet{Johnson2007}, but was considered as a CEMP-no star by \citet {Norris4}.  [Al/Fe] is mid to high, which means some operation of the Mg--Al cycle. This star  corresponds nicely to a class 4+. We note that  despite its s-elements, it presents
no enhancement of the heavy neutron--capture elements (of the 2$^{nd}$ and 3$ ^{rd}$ s-peak) as shown by  \citet{Norris4}, which is  consistent with our scenario. 

The star CS 22958-42 has a low [Ba/Fe], while [Sr/Fe] is slighty above the mean. [Mg/Fe] and [Al/Fe] are below the mean, while
[Na/Fe] is the highest in the sample;  the excess of Na with respect to Mg was noted by \citet{Allen2012}. In the [O/N] vs. [C/N] diagram, this object lies on the right side, but a bit low \citep{Maeder2015}, while it  
is in the main trend of the $^{12}$C/$^{13}$C vs. [C/N] plot. This star well corresponds to a class 4.

CS 29498-043 shares some  characteristics of the previous object.  [Sr/Fe]  and  [Ba/Fe] are close to the mean,        
 with more Mg and Al, thus providing some indications of the activity of  the Mg--Al cycle.  [Na/Fe] is marginally in excess.  The CNO triplet is low to mid. We also assign a class 4. 
The absence of  heavy neutron--capture elements was also emphasized, as for CS 22949-037 \citep{Norris4}.

We now turn to CS 30322-023 a star which shows a CNO triplet remarkably  corresponding to nearly pure CNO equilibrium (paper I). At the same time,
it presents a relatively high Ba content and was cataloged as "b-" by \citet{Allen2012}, this places it clearly in class 4++.  Both
[Mg/Fe] and [Na/Fe] are above the mean. The problem is that the  gravity of this star  is very low.
According to \citet{Masseron2006}
this star  may be a TP--AGB star, a point of view which we support.  Thus, in this case many of its properties could result from self--enrichment, nevertheless
according to these authors it shows no evidence of europium. We keep it as a CEMP--no star and assign a class 4++ to it.

HE 1012-1540 has mean values of [Sr/Fe] and [Ba/Fe]. It clearly shows
 strong evidences of the Ne--Na and Mg--Al cycles. 
The  [C/N] and [O/N] ratios are rather high,  clearly showing the signature of He-burning. Thus we assign a class  3 Na to this object.
In view of its Mg and Al abundances, it may not be far from a class 4.
 The  $^{12}$C/$^{13}$C  data, which would be most useful, are not available.

HE 1327--2326 was for some time the most iron--poor star known \citep{Frebel2005}, it has only a (high)  upper limit  [Ba/Fe] <1.46, thus it cannot formally by classified as a CEMP--no
star, although it shares the other characteristics of this class \citep{Norris4}. However, we have seen that stars showing some Sr and Ba  may belong to the  genetic family of CEMP--no stars, {\emph{i.e.} their abundances result from the same processes of mixing and mass loss as the other 
CEMP--no stars. The [Sr/Fe] ratio is the highest in the sample. The content in Na, Mg and  Al are also very high. At the same time, the CNO triplet is clearly low, thus providing  a dominant  signature of H-burning, while [(C+N+O)/Fe] is quite large.  Without any ambiguity, we assign  a class 4++
to it. We would say that HE 1327--2326 is an outstanding  illustration of 
our scenario for explaining the abundances of CEMP--no stars.

HE 1419-1324  shows a strong  excess of [Ba/Fe],  which in itself support  a "++" symbol. It  is recognized as a low-s star  \citep{Masseron2010,Allen2012}. There is no data about Sr and Na. [Mg/Fe] is close to the mean. [C/N] and [O/N] are below the mean,
while the ratio $^{12}$C/$^{13}$C is medium. It is likely a Class-4++ star, but the lack of data about Sr, Na, and Al makes the identification incomplete,  hence the symbol ":" .

HE 2331-7155 shows low-to-moderate values of [Sr/Fe] and [Ba/Fe]. The ratio [Na/Fe] is slightly lower than the median value.  The ratio  [Mg/Fe] is high, while [Al/Fe] is in the mean. The CNO triplet is low, especially [C/N]  = -1.23.  This is  a clear signature of  H-burning, which favors Class 2+ or 4. On the other hand, the moderate values of Sr and Ba and the high values of Mg indicate the possible previous contribution of He-burning. Thus,
we tend to assign Class 4 to this star.

The presently most  Fe-poor star known, SMSS 0313-6708, cannot be classified safely at present, since it essentially has upper bounds for its
very weak abundances of heavy elements. This is not the case  for Mg, and we have [Mg/Fe] > 3.3.  This extremely high value tends to imply a very heavy He-burning (labeled with ++). This is consistent with the relatively high upper bounds for [Sr/H] <-6.7 and [Ba/H]<-6.1 compared to [Fe/H]< -7.1, as obtained by \citet{Keller2014}.
Interestingly enough, in Paper I we could have both an upper bound  (-2.49)  and a lower bound (-3.09) for [(C+N+O)/H]. Adopting  the mean
[(C+N+O)/H] $\approx -2.8$,  we obtain a lower bound  [(C+N+O)/Fe] > 4.3  (which is used in Figs. \ref{CFE} and \ref{CNOSUM}).
The  lower limit  $^{12}$C/$^{13}$C $>$ 4.5  may allow some    H-burning. Therefore, this extreme object could belong to  Class 4++, but it is  uncertain in view 
of the difficulty of obtaining better data for this extreme object.

We note that the uncertain cases  essentially result from our having tried, rightly or wrongly,
to still classify a few stars with marginally insufficient data. 
There are only a few stars in the  group of Classes 2, compared to the groups of Classes 3 and 4. This may indicate that there are in general a few back-and-forth partial mixing motions between the H- and He-burning zones.
The main result is that the stars with sufficient data can generally be classified well in the proposed system.  The various classes  cover the observed variety of the chemical abundances of CEMP-no stars.
This supports the view that the effects of mixing (likely due  to rotation) and mass loss are responsible for the variety of composition in CEMP-no stars.

\subsection{Behavior as a function of [Fe/H]}

CEMP-no stars  span a range of more than  4.5 dex in [Fe/H] and [$\alpha$--elements/Fe].
(We have seen  in Paper I that the 
$\alpha$-elements, with an atomic mass number above 24 such as Ca and Si, behave like [Fe/H].)
We  now examine whether some trends appear  over this broad range of metallicities.

In each class, there are stars with a wide variety of [Fe/H] values. This is not 
 surprising since at each metallicity, there is probably a variety 
of rotation velocities and probably also of mass loss rates. Nevertheless, it seems that there is a  trend toward the  higher classes appearing at lower [Fe/H]. For Classes 2, the average [Fe/H]= -3.63;
for Classes 3, it is -3.72; and  for Classes 4,  the average is  -3.98. Certainly, more data are needed to ascertain  this trend based on a small sample,
which may be influenced by a few extreme values. 

The main criteria for CEMP stars are the [C/Fe] ratios, which we now examine. 
A plot of [C/Fe] vs.  [Fe/H]  for  the  stars  in Table 2 is shown in Fig. \ref{CFE}, which also indicates the standard deviation. Such an essential  plot was also shown by 
\citet{Norris4}. We see that there is a trend, but the relation is  poor. Without the four stars on the left, there would be 
no trend at all. The scatter around the average relation 
 is  large, amounting to about 1 dex.
We have a slightly better relation when we correlate  [(C+N+O)/Fe] with [Fe/H]  in Fig. \ref{CNOSUM}.

We may understand why the relation   of [(C+N+O)/Fe]  vs. [Fe/H] is better than for [C/Fe].  This may be because
the sum of C+N+O elements  only results from  the  number of  products (mainly C and O)  of He-burning that migrates into the H-burning region;  this amount is not 
modified by the CNO reactions in the source star. As to C,  its abundance is influenced by  the strength of both the mixing process and the CNO-burning
in the source stars.

In this respect, we note that the individual C, N, and O abundances may also be modified by the CNO processing in the CEMP-no stars themselves. The evolutionary effects of CNO processing on the carbon abundances have been been studied in detail as a function  of $\log g$ by \citet{Placco2014}, with the interesting result that it allows them to apply a correction on the C abundances of CEMP stars. This leads to a much better estimate of the fraction of these stars as a function of [Fe/H]. We may point out that the effects studied by \citet{Placco2014}  bring essentially no  change for the sum  [(C+N)/Fe]
 (or [(C+N+O)/Fe]), while they evidently change [C/Fe].

The  relation between  [(C+N+O)/Fe]  and  [Fe/H]  may suggest that this is essentially the higher strength of the 
mixing (and maybe the higher rotation) in the source stars at lower [Fe/H], which is responsible for the peculiarities of the CEMP-no stars,
where the exchanges between the two burning regions are more
intense at lower [Fe/H]. This higher effect of mixing at lower Z was first found (and explained)   in the context of  stellar models at  SMC composition \citep{Maeder2001}. It was then confirmed by studying the origin of primary nitrogen in low-metallicity galaxies  \citep{MM2002}, after the evidence of primary nitrogen had  firs beent  been demonstrated by \citet{Edmunds1978}.
That we observe strong mixing effects in  CEMP-no stars indicates that  mixing continues to grow at much lower metallicities 
than for   the SMC.

\section{Conclusions}
 The classes we proposed on the basis of  the nucleosynthesis in successive phases of mixing between the H- and He-burning regions cover the great variety of C, N, O, Na, Mg, and Al abundances observed in CEMP-no stars well. The classification is essentially model independent.
A few transition and uncertain  cases  appear mainly by insufficient data, but there is no  dichotomy between the classes considered and the variety of abundances observed. 
 A possible extension of this classification to the other types of CEMP stars with [Fe/H]  higher than $\approx$ -2.5 could be envisaged in future.

We may wonder at what stage in the evolution of the massive stars does occur the matter ejection leading to CEMP--no stars. There are  four stars
in Classes 2, 17 in Classes 3 and 9 in Classes 4. The total of 13 stars ejected in Classes 2 and 4 show the signatures of CNO processing, with the
last nuclear reactions having occurred before the ejection (with more turnovers in Classes 4 than in 2).  This means that the H-burning shell was still active in the outer stellar layers just before the ejection. In terms of evolution, this points clearly toward ejection during the
supergiant stages, whether blue or red.

The situation is different for the 17 stars in Classes 3. There, the main nuclear signature before the ejection is that of  He-burning.
This necessarily implies that the H-burning shell was nearly or totally extinct at the time of large mass ejections. This  usually occurs 
very late in the evolution, shortly  before the supernova explosion.
This may correspond to the  suggestion by \citet{Moriya2015} that very intense mass ejections due to pulsational instabilities 
occur in the red supergiant stage
of metal-free stars shortly before the supernova explosions, thus creating very dense circumstellar shells. Another possibility is that of mass ejection
at various stages for homogeneous evolution due to  
mixing, a situation which may occur in very fast-rotating massive stars  \citep{Maeder1987}. 

On the whole, this classification, along with the tests we performed in Paper I support the view that rotating massive stars with mixing and mass
loss  strongly influenced  the early chemical  (and probably spectral) evolution of galaxies.\\

\small{Acknowledgments: We express our deepest thanks to our friends and colleagues Corinne Charbonnel, Patrick Eggenberger, Sylvia Ekstroem, and Arthur Choplin for very useful discussions. We also thank  Prof. Timothy Beers for providing  new data in advance of publication and for very helpful remarks.}

\appendix
\section{CEMP-no stars that cannot be classified}
There are 16 stars that we cannot  classify because some essential chemical abundances are not available. They deserve some brief comments.
For CS 22166-016, the values of [Ba/Fe] and  [Sr/Fe] show no excesses, the same for [Na/Fe] and [Mg/Fe]. These values imply that it is not in
Class 4, 4+, 4++, 3+, or 3++. However, the lack of the CNO triplet prevents us from deciding whether it belongs to Class 2 or 3.
For CS 22878-027, there is no evidence of significant excess of s-elements, and the absence of data for the CNO triplet also prevents the classification (only a non--constraining lower limit for [C/N] is available).
 CS 22885--096 has a low [C/Fe]= 0.60 \citep{Norris2001}, which is a marginally  low value for a CEMP-no star. It is, however, considered to be 
a CEMP-no star by \citet{Allen2012}. The lack of the CNO triplet and of [Na/Fe] does not allow us to assign a class to this star, which has extremely low 
values of [Sr/Fe] and [Ba/Fe].
CS 22956--028 is a binary star discussed by \citet{Sneden2003}, and it is considered to be a low-s star by \citet{Masseron2010}. 
The [Ba/Fe] ratio of 0.16 is above the mean, but it  is not a clear excess according to Fig. \ref{BA}.
The lack of the 
Sr- and Na-abundances  prevents any reliable classification.
 CS 22957--027 shows radial velocity variations \citep{Preston2001,Norris4}, it is probably in  Class 2 in view of the ratio $^{12}$C/$^{13}$C, but 
it cannot be confirmed owing to 
the lack of O- and Na-data.
CS 30314-067  has  values of [Sr/Fe] close to the mean and  of [Ba/Fe] slightly below the mean. [Mg/Fe] and  [Al/Fe] are in the mid range, while [Na/Fe] is much lower than the mean.  There is no clear evidence of  the Ne--Na
and Mg--Al cycles. The low value of [C/N]  is  a signature of efficient H-burning.  We would place it in Class 4, however the lack of [O/N] and $^{12}$C/$^{13}$C prevents any
 confirmation and leaves open the possibility of a Class 2+.
For G 77--61, the high upper limit for [Ba/Fe] is not  constraining and does not allow us to determine a class for this star.

For HE 0007-1832, the ratio [Ba/Fe] is at the limit for showing an excess, [C/N] is well above the mean, however 
 the available data are clearly insufficient to allow a classification.
HE 0146-1548 has a value of [Ba/Fe]  below the mean and [Sr/Fe] is  close to the mean. [Mg/Fe] is above the mean, and there is some marginal indications of the Ne--Na and Mg--Al cycles.  These properties, together with the low  $^{12}$C/$^{13}$C ratio, correspond to a Class 4 or 4+,
however the absence of values for [C/N] and [O/N] prevents a safe assignation.
 For  HE 1150-0428, [Ba/Fe] has a mean value, while [Sr/Fe]
is  slightly above the mean, thus there is no evidence of significant enhancements in s--elements. [Mg/Fe] is below the mean. There is no information about
 the Ne--Na and Mg--Al cycles. From the low $^{12}$C/$^{13}$C and [C/N] ratios, we would propose a Class 2+ or 4, but we have not enough data to decide. 
For  HE 1300-0641, a lot of data is missing and even a tentative classification is impossible.
 We can only say that there is no indication of excesses for s-elements. [Mg/Fe] is 
low, and in view of [Al/Fe], the Mg--Al cycle has certainly not operated. HE 1300-2201 presents a significant enrichment in [Sr/Fe], but  [Ba/Fe] is only slightly above the mean. 
The value of [Mg/Fe] is below the mean, and there is no evidence of the Mg--Al cycle. Although this star has no more data than the previous one,
the significant [Sr/Fe]  excess is somehow constraining. The class is probably 3+, nevertheless the lack of the
CNO triplet and of [Na/Fe] prevents a reliable classification.
HE  1330-0354 shows no Ba excess, but [Sr/Fe] is just above the limit of what we consider as an excess.   It has little Mg and Al. Thus, it could  be in class 3+, however the lack of data makes this possibility unconfirmed. HE 2142-5656 does not show enhancements of
s--elements, neither of the occurrence of Ne--Na and  Mg--Al cycles. All values are not far from the means.  The same is true for [C/N], however   the lack of [O/N] and $^{12}$C/$^{13}$C ratios prevents an unambiguous classification, 
HE 2202-4831 also shows no evidence of excesses of s-elements, the values of [Ba/Fe] and [Sr/Fe] being clearly below the means. Mg is below the mean, while [Na/Fe] is above the mean, close to the limit of having an excess. The absence of data on the CNO triplet prevents us to assign a class. 
For HE 2247-7400, the situation is about the same: no sign of excess of s-elements, both Mg and Na below the mean. Here also, the absence 
of the CNO triplet does not allow a classification.

\bibliographystyle{aa}
\bibliography{Maeder-CEMP}
\end{document}